\documentstyle[floats,prd,aps,epsfig,eqsecnum,12pt]{revtex}
\makeatletter
\newbox\tempboxa
\newdimen\captionboxsubcount
\def\capsize#1{\captionboxsubcount=#1pt}
\newdimen\captionboxsub
\captionboxsub=\hsize \advance\captionboxsub by -\captionboxsubcount
\advance\captionboxsub by -\captionboxsubcount
\long
\def\@makecaption#1#2{
\setbox\@tempboxa\hbox{#1 #2}
\ifdim \wd\@tempboxa >\captionboxsub
\rightskip=\captionboxsubcount \leftskip=\captionboxsubcount #1 #2
\else \hbox to\hsize{\hfil\box\@tempboxa\hfil}
\fi}
\makeatother
\capsize{30}

\begin{document}

\begin{titlepage}
\begin{flushright}
\begin{minipage}{5cm}
\begin{flushleft}
\small
\baselineskip = 13pt
YCTP-P04-99\\ SU--4240--694\\ hep-ph/9903359 \\
\end{flushleft}
\end{minipage}
\end{flushright}
\begin{center}
\Large\bf
Chiral Phase Transition for
\mbox{\boldmath${SU(N)}$} Gauge
Theories via an Effective Lagrangian Approach.
\end{center}
\vfil
\footnotesep = 12pt
\begin{center}
\large
Francesco {\sc Sannino}$^a$ \footnote{ Electronic address : {\tt
francesco.sannino@yale.edu}}\quad and \quad
 Joseph {\sc Schechter}$^b$
\footnote{ Electronic address : {\tt
schechte@suhep.phy.syr.edu}}\\

{\it $^a$ Department of Physics, Yale University, New Haven,~CT~06520-8120,~USA.}\\ \vskip .5cm

{\it $^b$ Department of Physics, Syracuse University,
Syracuse,~NY~13244-1130,~USA.}
\end{center}
\vfill
\begin{center}
\bf
Abstract
\end{center}
\begin{abstract}
\baselineskip = 17pt
We study the chiral phase transition for vector-like $SU(N)$ gauge
theories as a function of the number of quark flavors $N_f$ by making
use of an anomaly-induced effective potential. We modify an
effective potential of a previous work, suggested for $N_f < N$,
and apply it to larger values of $N_f$ where the phase transition
is expected to occur. The new effective potential depends
explicitly on the full $\beta$-function and the anomalous dimension
$\gamma$ of the quark mass operator. By using this potential we
argue that chiral symmetry is restored for $\gamma <1$. A
perturbative computation of $\gamma$ then leads to an estimate of
the critical value $ N_f^c$ for the transition.
\end{abstract}
\begin{flushleft}
\end{flushleft}
\vfill
\end{titlepage}

\section{Introduction}

The phase structure of strongly coupled gauge field theories as a function
of the number of matter fields $N_f$ is a problem of general interest. Much
has been learned about the phases of supersymmetric theories in recent years
\cite{Seiberg,Seiberg-Witten,IntSeiberg,Peskin,DiVecchia}. An equally
interesting problem is the phase structure of a non-supersymmetric $SU(N)$
theory as a function of the number of fermion fields $N_f$. At low enough
values of $N_f$, the chiral symmetry $SU(N_{f})_L \times SU(N_{f})_R$ is
expected to break to the diagonal subgroup. At some value of $N_f$ less than
$11N/2$ (where asymptotic freedom is lost), there will be a phase transition
to a chirally symmetric phase. Whether the transition takes place at a
relatively small value of $N_f$ \cite{mawhinney} or a larger value remains
unknown. The larger value ($N_f / N \approx 4$) is suggested by studies of
the renormalization group improved gap equation \cite{ATW} and is associated
with the existence of an infrared fixed point. A recent analysis \cite{ASe}
indicates that instanton effects could also trigger chiral symmetry breaking
at comparably large value of $N_f/N$. Besides being of theoretical interest,
the physics of a chiral transition could have consequences for electroweak
symmetry breaking \cite{AS}, since near-critical gauge theories provide a
natural framework for walking technicolor theories \cite{ATW2}.

If a phase transition is second order, a useful approach is to find a
tractable model in the same universality class. For chiral symmetry, a
natural order parameter is the $N_f\times N_f$ complex matrix field $M$
describing mesonic degrees of freedom. If the meson degrees of freedom are
the only ones that develop large correlation lengths at the phase
transition, then the transition may be studied using an effective
Landau-Ginzburg theory. This time-honored approach has been used, for
example, to study the QCD finite temperature transition for $N_f = 2$ \cite
{PisWil}.

For the zero-temperature transition as a function of $N_f$, a similar
approach might also be tried. It was suggested in Ref.\cite{ATW}, however,
that while the order parameter vanishes continuously as $N_f \rightarrow
N_f^c$, the transition is not second order. With the gap equation dominated
by an infrared fixed point of the gauge theory, the transition was argued to
be continuous but infinite order. It has also been noted \cite{sekhar} that
because of the associated long range conformal symmetry, the masses of all
the physical states, not just the scalar mesons are expected to scale to
zero with the order parameter.

In this paper we nevertheless suggest that an effective potential using only
the low lying mesonic degrees of freedom might be employed to model at least
some aspects of the zero-temperature chiral phase transition. The key
ingredient is the presence of a new non-analytic potential term that emerges
naturally once the anomaly structure of the theory is considered. The
anomalies also provide a link between this effective potential term and the
underlying gauge theory.

To deduce the anomaly induced effective potential we modify an effective
potential \cite{toy,HSS,Sannino} developed for $N_f< N$, and apply it to the
range $N_f > N$. The effective potential of Refs. \cite{toy,HSS,Sannino} was
suggested by starting with the effective Lagrangian for super-QCD and
considering how the gluinos and squarks decouple below a supersymmetric
breaking scale $m_s$. In Reference \cite{HSS}, it was noted that this
potential can also be constructed, once the trace and axial anomaly
constraints are saturated at one loop, by assuming holomorphicity.

In Section II, we set the stage by providing a brief review of the SUSY QCD
effective potential for $N_f < N$, and comparing it to the one-loop,
anomaly-induced effective QCD potential of Refs. \cite{toy,HSS,Sannino}. In
Section III an effective potential valid to all orders in the loop
expansion, and appropriate for the range $N_{f} > N$, is proposed. It
utilizes only mesonic variables to capture the low energy dynamics. In
Section IV we use this potential to discuss the zero-temperature phases of
an $SU(N)$ gauge theory as a function of $N_f$. We use the singular behavior
of the curvature of the effective potential at the origin as a signal for
chiral restoration. Assuming that the transition is governed by an infrared
fixed point of the theory, we deduce that chiral symmetry is restored,
together with long-range conformal symmetry, when $\gamma <1$, where $\gamma$
is the anomalous dimension of the mass operator. Finally we note that by
using the perturbative expansion of $\gamma $, chiral symmetry is predicted
to be restored above $N_f^c \approx 4N $, in agreement with a gap equation
analysis. In Section V we summarize our results and provide some discussion.
In Appendix A we examine some higher loop effects in the effective potential.

\section{Review of the Anomaly-Induced Effective Potential}

We start by recalling the role of the effective potential in super QCD
theories. For $N_f<N$, the effective low energy superpotential takes the
Affleck-Dine-Seiberg (ADS) \cite{refADS} form
\begin{equation}
W_{ADS}(T)=-C_s\left( N,N_f\right) \left[ \frac{\Lambda _S^{3N-N_f}}{{\rm %
det }T}\right]^{\frac 1{N-N_f}}\ ,  \label{ads}
\end{equation}
where the composite meson superfield $T$ has the same quantum numbers as $Q%
\tilde{Q}$, with $Q$ and $\tilde{Q}$ being the quark superfields, and $%
\Lambda _S$ is the intrinsic scale of super QCD (SQCD). In this
instanton-generated super potential, the exponent of $\Lambda_S$ is the
coefficient of the lowest order term in the supersymmetric $\beta$ function.
Through a suitable decoupling procedure, one can show \cite{Seiberg} that
the function $C_s\left( N,N_f\right)$ takes the form $C_s\propto \left(
N-N_f\right) K^{1/(N-N_f)}$ where $K$ is an arbitrary constant independent
of the number of colors and flavors. By an explicit instanton calculation
one finds $K=1$. In supersymmetric theories the axial anomaly together with
the superconformal anomaly can be cast in the same chiral supermultiplet.
This fact together with the holomorphic constraint has led to the idea that
the ADS superpotential can be constructed by using only the information
contained at one-loop in the underlying theory.

According to the ADS potential there is no stable vacuum in the massless
theory for any $N_f<N$. Furthermore, the ADS superpotential is singular for $%
N_f=N$. Seiberg argued that the superpotential should be modified for $N_f=N$
and that the singularity signals the occurrence of new, massless degrees of
freedom. In the case $N_f=N$, these massless degrees of freedom are
identified with the superfield baryon $B\propto \epsilon
^{c_1,...,c_N}\epsilon _{i_1,...,i_{N_f}}Q_{c_1}^{i_1}\cdots
Q_{c_N}^{i_{N_F}}$ (a similar construction holds for the $\tilde{B}$ field).
Depending on the choice of the vacuum, chiral symmetry can be either broken
or unbroken. For $N_f > N$, a variety of phases is possible depending on $%
N_f $ \cite{Seiberg}.

In Refs.~{\cite{toy,HSS}}, an attempt was made to construct a potential in
the same spirit as the ADS superpotential for a (non-SUSY) SU(N) gauge
theory by saturating at one-loop the energy-momentum-trace and axial
anomalies and imposing holomorphicity. The result, for $N_f < N$, was
\begin{equation}
V=-C\left( N,N_f\right) \left[ \frac{\Lambda ^{\frac{11}3N-\frac 23N_f}}{%
{\rm det}M}\right] ^{\frac{12}{11\left( N-N_f\right) }}+{\rm h.c.},
\label{hss}
\end{equation}
where $M_i^j $ is the $N_f\times N_f$ complex matrix field possessing the
same quantum numbers as $q_i\tilde{q}^j$. Upon quantization, it would
describe mesonic degrees of freedom. $C\left( N,N_f\right) $ is a
coefficient which, after defining a suitable one loop decoupling procedure,
turns out to be proportional to $\left( N-N_f\right) D(N)^{1/(N-N_f)}$ where
$D(N)$ is an unknown function of N. $\Lambda $ was taken to be the
confinement scale of the theory and its exponent in Eq.~(\ref{hss}) is the
first coefficient in the perturbative expansion of the $\beta $ function.
For $N_f < N$ this potential displays a fall to the origin.

In Ref.~\cite{toy} it was noted that the fall to the origin can be cured by
introducing non holomorphic terms that implement spontaneous chiral symmetry
breaking. The non-holomorphic piece is constructed so that it does not
contribute to either the $U(1)_A$ anomaly or the trace anomaly. The
holomorphic piece was shown to play a special role in that it alone
describes the $\eta^{\prime}$ self interactions including an $\eta^{\prime}$
mass term. For $N_f < N$, \cite{HSS}
\begin{equation}
M^2_{\eta^{\prime}}\propto \frac{N_f} {N-N_f}\Lambda^2 .  \label{eta}
\end{equation}
The large $N$ behavior was anticipated in Ref.\cite{Wittena,Veneziano}. The
mass squared of the $\eta^{\prime}$-field is seen to diverge as $N_f
\rightarrow N$, a singularity also present in $V$ (Eq.~(\ref{hss})). In the
analogous supersymmetric case, the corresponding singularity in Eq.~(\ref
{ads}) is overcome by the appearance of additional baryonic light degrees of
freedom. However, there is no indication that this is the case for a
QCD-like theory so the singularity as $N_f\rightarrow N$ is very likely an
artifact of the perturbative approximations leading to Eq.~(\ref{hss}).

\section{A Nonperturbative Effective Potential}

In this section we construct an effective potential valid to all orders in
the loop expansion and appropriate for the range $N_{f}>N$. The new
ingredients are:

\begin{itemize}
\item[i)]  {Using the full, rather than the one loop, beta function in the
trace anomaly saturation.}

\item[ii)]  {Taking account of the anomalous dimension of the fermion mass
operator.}
\end{itemize}

This anomaly-induced effective potential is based on the QCD trace and $%
U_A(1)$ anomalies:
\begin{eqnarray}
\theta _m^m &=&\frac{\beta (g)}{2g}F_a^{mn}F_{mn;a}\equiv 2bH\ , \\
\delta _{U_A(1)}{\ V_{QCD}} &=&N_f\alpha \frac{g^2}{32\pi ^2}\epsilon
_{mnrs}F_a^{mn}F_a^{rs}\equiv 4N_f\alpha G\ ,  \label{anomalies}
\end{eqnarray}
where we have defined $\displaystyle{\beta (g)\equiv- b g^3 /(16\pi ^2)}$.
We take the coupling to be defined at some low energy scale appropriate for
the phase transition to be studied. Eventually, we will assume that the
transition is governed by an infrared fixed point of the gauge theory and
set $b= 0$. At one loop, $\displaystyle{b=\frac{11}3N-\frac 23N_f}$.

$H$ and $G$ are composite fields describing, upon quantization, scalar and
pseudoscalar glueballs \cite{joe}. The general, non derivative effective
potential saturating the anomalies is:
\begin{equation}
V=-F\sum_{n}c_n{\rm ln} \left( \frac{{\cal O}_n}{\Lambda ^{d_n}}\right) +
{\rm h.c.}\ ,  \label{general}
\end{equation}
where $\Lambda$ is some fixed intrinsic scale of the theory and where ${F}
=H+i\delta G$ with $\delta$ a positive constant (a negative $\delta$ is
equivalent to interchanging $F$ with $F^{\dagger}$) to be chosen. The ${\cal %
O}_n$ are gauge invariant fields built out of the relevant degrees of
freedom with naive mass dimension $d_n$ and axial charge $q_n$. The presence
of the ${\rm ln}({\cal O}_n/\Lambda ^{d_n})$ structure insures the correct
implementation of the underlying anomalies at the effective potential level.

The anomaly constraints are:
\begin{equation}
\sum_{n}c_{n} D_{n} =b\ ,\qquad \sum_{n}c_{n}q_{n}=\frac{2N_f}{\delta },
\label{sum}
\end{equation}
where $D_n = d_n - \gamma_n$ is the dynamical, scaling dimension of ${\cal O}%
_n$, with $\gamma_n$ the anomalous dimension. We remark that the derivation
of V is based on making explicit scale and $U(1)_A$ transformations on the
composite operators ${\cal O}_n$. Since the scaling dimension enters as a
parameter in this approach it is natural to associate it with the dynamical
quantity $D_n$.

Here we choose $\delta=1$ \cite{KZ}, but we will note in Section IV that our
result is independent of the specific positive value assigned to $\delta$.
The gauge degrees of freedom described by the dimension-four field $F$, have
been introduced as an intermediate device to implement correctly at the
effective potential level the underlying anomalous transformations. The
anomalous dimension of $F$ is zero. As indicated in the introduction, we
will build the potential out of the $N_f\times N_f$ complex meson matrix $%
M_i^j$ transforming as the operator $q_i\tilde{q}^j$. So we assign naive
mass dimension 3 to $M_i^j$. The operator $q\tilde{q}$ acquires an anomalous
dimension $\gamma $ when quantum corrections are considered and the full
dynamical dimension is thus $3-\gamma$. To make our effective potential
capture the low-energy quantum dynamics of the underlying theory, we take $%
3-\gamma$ to be the scaling dimension of $M_i^j$. The anomalous dimension $%
\gamma$ is of course a function of the coupling $g$, which in turn depends
on the relevant scale.

We next make the simplifying assumption that the fields ${\cal O}_{n}$ in
the potential Eq.~(\ref{general}) may be restricted to a minimal set (with
lowest dimension) sufficient to satisfy the anomaly constraints Eq. (\ref
{sum}). Thus we include only two terms ${\cal O}_{1}=-F$ and ${\cal O}_{2}=%
{\rm det}M$. Including additional terms would introduce arbitrary parameters
in the model, which seems inappropriate for an initial investigation.
Retaining just the minimal set is plausible (see section VII of Ref.~{\cite
{GJJS}}) and would correspond to the ''holonomic'' structure which emerges
if the potential is considered to arise (\cite{toy,HSS,Sannino}) from broken
super QCD. In the same spirit we take det$M$ to have the scaling dimension $%
(3-\gamma )N_{f}$.

%This is, in effect, an assumption of holomorphy, meaning a
%potential of the form $\chi(\Phi) + {\rm h.c}.$ where $\chi$ is a
%function of the generic complex field $\Phi$. The holomorphy
%constraint is exact in supersymmetric theories, while in QCD-like
%theories is an assumption which seems to provide interesting
%results consistent with the standard lore \cite {GJJS}. Finally we
%make the asumption that the quantum scaling behavior of the
%underlying theory is captured completely by the fields $M_i^j$.
%Products of these fields will be taken to scale as simple multiples
%of the $M_i^j$. Thus, for example, we take ${\rm det} M $ to have
%scaling dimension $% (3-\gamma)N_f$.

The potential in Eq.~(\ref{general}) then takes the form
\begin{equation}
V\left(F,M\right)=\left(\frac{\beta(g)}{g^3}16\pi^2 + (3-\gamma)N_f \right)
\frac{F }{4} {\rm ln} \left(\frac{-F}{\Lambda^4}\right) - F {\rm ln} \left(
\frac{{\rm det} M} {\Lambda^{3 N_f}} \right) + AF + {\rm h.c.} \ ,
\label{assqcd}
\end{equation}
where $A$ is a dimensionless constant that cannot be fixed by saturating the
QCD anomalies and assuming holomorphicity. In fact, the term $AF$ does not
contribute to $\theta^m_m$ and is also a chiral singlet. The potential of
Eq.~(\ref{assqcd}) is seen to be consistent with the constraints in Eq.~(\ref
{sum}) when $\displaystyle{4c_1=b-(3-\gamma)N_f} $ and $\displaystyle{c_{2}=
1}$. The coupling $g$ and anomalous dimension $\gamma$ are defined at some
scale $\mu$. For our study of a chiral phase transition governed by an
infrared stable fixed point, $g$ will be the fixed-point coupling and $%
\gamma $ will be the associated anomalous dimension. The $\beta$ function
will then vanish.

To construct a potential depending only on the meson degrees of freedom we
''integrate out'' the gluonic degrees of freedom by imposing the field
equation $\displaystyle{\partial V/\partial F=0}$, which provides
\begin{equation}
F=-e^{\frac{4A}{f(g)}-1} \Lambda^{4} \left[ \frac{\Lambda ^{3N_f}}{{\rm det}%
M }\right] ^{\frac 4{f(g)}}\ ,
\end{equation}
where
\begin{equation}
f(g)=-\frac{\beta (g)}{g^3}16\pi ^2-(3-\gamma )N_f \ .  \label{function}
\end{equation}
After substituting the expression for $F$ in Eq.~(\ref{assqcd}) we obtain:
\begin{equation}
V=-C \Lambda^{4}\left[ \frac{\Lambda ^{3N_f}} {{\rm det}M}\right] ^ {\frac 4{%
f(g)}}+{\rm h.c.} \ ,  \label{ass}
\end{equation}
where $C$ is related to $A$ via:
\begin{equation}
C=\frac {f(g)}{4e}\exp \left[ {\frac{4A} {f(g)}}\right] \ .
\label{coefficient}
\end{equation}
Finally we integrate out the $\eta^{\prime}$ field, which can be isolated by
setting
\begin{equation}
{\rm det}M=|{\rm det}M|e^{i\phi }\ ,  \label{prime}
\end{equation}
where $\phi \propto \eta ^{\prime }$. This is done anticipating that the $%
\eta^{\prime}$ will be heavy with respect to the intrinsic scale of the
theory and the other mesonic degrees of freedom. Now using Eq.~(\ref{prime})
we derive the field equation $\phi=0$ which leads to the final potential
%The $\eta ^{\prime }$
%is integrated out along with the non-mesonic degrees of freedeom
%since we want to build the effective potential from only the
%mesonic degrees of freedom associated with the chiral phase
%transition. Using Eq.~(\ref{prime}) we derive the field equation
%$\phi=0$ which leads to the final potential
\begin{equation}
V=-2C \Lambda^{4}\left[\frac{\Lambda^{3N_f}} {|{\rm det} M|}\right]^{\frac{4%
}{f(g)}} \ .  \label{inter}
\end{equation}
The shape of this potential is determined by the function $f(g)$ Eq.~(\ref
{function}). The potential of Eq.(\ref{hss}) simply used the lowest order
perturbative expansion of $f(g)$ ($\gamma =0$ and $\displaystyle {-\frac{%
\beta (g)}{g^3}16\pi^2 =\frac{11}3N-\frac 23N_f}$).

Our interest here is in the range $N < N_f < (11/2)N$ where the chiral phase
transition is expected to occur. For $N_f$ close to $(11/2)N$, a weak
infrared fixed point will occur. The $\beta$ function will be negative and
small at all scales and $\gamma$ will also be small. Thus $f(g)$ will be
negative. As $N_f$ is reduced, the fixed point coupling increases as does $%
\gamma$. We will argue (in Appendix A), however, that in the range of
interest, $f(g)$ will remain negative ($(3-\gamma)N_f >
-(\beta(g)/g^{3})16\pi^2$). The potential in Eq.~(\ref{inter}) may then be
written as
\begin{equation}
V=+2|C|\Lambda^{4}\left[ \frac{|{\rm det}M|} {\Lambda ^{3N_f}} \right]^{%
\frac 4{\frac{\beta (g)} {g^3}16\pi ^2+(3-\gamma )N_f}}\ .  \label{grandeNf}
\end{equation}
It is positive definite and vanishes with the field $|{\rm det}M|$.

\section{The Chiral Phase Transition}

To study the chiral phase transition, we need the combined effective
potential
\begin{equation}
V_{tot}=V + V_I \,
\end{equation}
where $V_I$ is a generic potential term not associated with the anomalies.
It is instructive, however, to investigate first the extremum properties of
the anomaly term (Eq.~(\ref{grandeNf})). Assuming the standard pattern for
chiral symmetry breaking $SU_R(N_f)\times SU_L (N_f)\rightarrow SU_V (N_f)$
, $M^i_j$ may be taken to be the order parameter for the transition. For
purposes of this discussion, we restrict attention to the vacuum value of $%
M^i_j$, which can be rotated into the form
\begin{equation}
M^i_j=\delta^i_j \rho \ ,  \label{vacuum}
\end{equation}
where $\rho \geq 0$ is the modulus. Substituting (\ref{vacuum}) in the
anomaly induced effective potential gives the following expression:
\begin{equation}
V=+2|C|\Lambda^{4} \left[\frac{\rho}{\Lambda^{3}} \right]^{\frac{4N_f}{\frac{%
\beta(g)}{g^3}16 \pi^2 + (3 - \gamma)N_f }} \ .  \label{rhopotential}
\end{equation}

Recall that ($(3-\gamma)N_f > -(\beta(g)/g^{3})16\pi^2$) in the range of
interest. The first derivative $\partial V/\partial \rho$ vanishes at $\rho=0
$ provided that
\begin{equation}
\frac{4N_f}{\frac{\beta (g)}{g^3}16\pi ^2+(3-\gamma )N_f}>1,
\label{1stcondition}
\end{equation}
a condition that is clearly satisfied. The second derivative,
\begin{equation}
\frac{\partial ^2V}{\partial \rho ^2}\propto \rho ^{\left[\frac{4N_f}{\frac{
\beta(g)} {g^3}16\pi ^2+(3-\gamma )N_f}-2 \right]} \ ,  \label{2nd}
\end{equation}
also vanishes at $\rho =0$ if the exponent in Eq.~(\ref{2nd}) is positive.
The second derivative at $\rho =0$ is a positive constant when the exponent
vanishes, and it is $+ \infty$ for
\begin{equation}
\frac{4N_f}{\frac{\beta (g)}{g^3}16\pi ^2+(3-\gamma )N_f}-2<0.
\label{criticality}
\end{equation}
The curvature of $V_{tot}$ at the origin is given by the sum of the two
terms $\frac{\partial ^2V}{\partial \rho ^2}$ and $\frac{\partial^2V_I}{%
\partial \rho ^2}$, evaluated at $\rho = 0$.

To proceed further, we assume that the phase transition is governed by an
infrared stable fixed point of the gauge theory. We thus set $\beta (g)=0$.
The curvature of V at the origin is then $0$ for $\gamma >1$, finite and
positive for $\gamma =1$, and $+\infty $ for $\gamma <1$. The value of $%
\gamma $ depends on the fixed point coupling, which in turn depends on $N_{f}
$. As $N_{f}$ is reduced from $(11/2)N$, the fixed point coupling increases
from $0$, as does $\gamma $. Assuming that $\gamma $ remains monotonic in $%
N_{f}$, growing to $1$ and beyond as $N_{f}$ decreases, there will be some
critical value $N_{f}^{c}$ below which $\frac{\partial ^{2}V}{\partial \rho
^{2}}$ vanishes at the origin. The curvature of $V_{tot}$ will then be
dominated by the curvature of $V_{I}$ at the origin. For $N_{f}=N_{f}^{c}$,
there will be a finite positive contribution to the curvature from the
anomaly-induced potential. For $N_{f}>N_{f}^{c}$ ($\gamma <1$), $V$
possesses an infinite positive curvature at the origin, suggesting that
chiral symmetry is necessarily restored. We will here take the condition $%
\gamma =1$ to mark the boundary between the broken and symmetric phases, and
explore its consequences. This condition was suggested in Ref. \cite{CG},
based on other considerations. It is straightforward to see that here, this
condition is independent of the value assigned to $\delta $ in Eq.~(\ref{sum}%
). The $\delta $ parameter enters the potential multiplying $\beta (g)$ and
is therefore irrelevant when $\beta (g)=0$.

We next investigate the behavior of the theory near the transition by
combining the above behavior with a simple model of the additional,
non-anomalous potential $V_I$. We continue to focus only on the modulus $%
\rho $ and take the potential to be a traditional Ginzburg-Landau mass term,
with the squared mass changing from positive to negative as $\gamma - 1$
goes from negative to positive:
\begin{equation}
\left( 1 - \gamma \right) \Lambda ^{-2}\rho ^2 \ .  \label{pot2}
\end{equation}
Additional, stabilizing terms, such as a $\rho^4$ term, could be added but
will not affect the qualitative conclusions. The full potential is then
\begin{equation}
V_{tot}=2|C|\Lambda^{4}(\frac{\rho}{\Lambda^3})^{\frac{4}{3-\gamma}} -
\left(\gamma - 1\right)\Lambda ^{-2} \rho^2 \ .  \label{potenzialetto}
\end{equation}
{}For $\gamma >1$ (but $<3$), the first term stabilizes the potential for
large $\rho$, and the potential is minimized at
\begin{equation}
< \rho >=\Lambda^{3} \left[\frac{(\gamma-1)(3-\gamma)}{4|C|}\right]^{\frac{1
}{2} \frac{3-\gamma}{\gamma-1}} \ ,  \label{exponent}
\end{equation}
It seems to us that this form may very well represent a generic extension of
the Landau- Ginzburg potential to the present case.

In the limit $\gamma \rightarrow 1$ this expression reduces to
\begin{equation}
<\rho >=\Lambda^3 \left[\frac{\gamma - 1} {2|C|}\right]^{\frac{1}{\gamma-1}}
\,  \label{lexponent}
\end{equation}
which describes an infinite order phase transition as $\gamma \rightarrow 1$
, in qualitative agreement with the gap equation studies. This behavior
would not be changed by the addition of higher power terms ($\rho^4, \rho^6,
...$) to the potential.

It is also interesting to describe how the order parameter $\rho $
approaches zero at the critical point (i.e. $\gamma =1$) as a function of
the quark mass. At the effective potential level (for $m\ll \Lambda $) the
quark mass enters in the following way
\begin{equation}
-m{\rm Tr}\,\left[ M+M^{\dagger }\right] =-2N_{f}m\rho ,
\end{equation}
where $m$ is a diagonal quark mass. This new operator when added to the
potential in Eq.~(\ref{potenzialetto}) yields
\begin{equation}
<\rho >_{\gamma =1}=\frac{m\Lambda ^{2}N_{f}}{2|C|}\ .
\end{equation}

The curvature of the potential Eq.~(\ref{potenzialetto}) at the minimum
describes a mass associated with the field $\rho $. To interpret this mass
physically, one should construct the kinetic energy term associated with
this field (at least to determine its behavior as a function of $\gamma -1$%
). We hence rescale $\rho $ to a field $\sigma $ via $\displaystyle{\rho
=\sigma ^{3-\gamma }\Lambda ^{\gamma }}$ with $\sigma $ possessing a
conventional kinetic term $-\frac{1}{2}(\partial ^{\mu }\sigma )^{2}$%
\footnote{%
The present rescaling procedure is consistent with the respect to the
construction of the energy momentum tensor at the effective potential level.
\par
It is interesting to notice that in the variable $\sigma $ the effective
potential reads
\par
\[
V=2|C|\sigma ^{4}-(\gamma -1)\sigma ^{2(3-\gamma )}\Lambda ^{2(\gamma -1)}.
\]
\par
Clearly the first term is conformally invariant and the second term can be
understood as a small deviation from conformality. We expect this effective
potential to be a suitable generalization of the Ginzburg-Landau theory when
a global symmetry (associated with the non vanishing vacuum expectation
value of the order parameter $\sigma $) is restored together with the
conformal symmetry. }. This then leads to the following result for the
physical mass $M_{\sigma }$ and $<\sigma >$
\begin{equation}
<\sigma >\simeq \left[ \frac{\gamma -1}{2|C|}\right] ^{\frac{1}{2(\gamma -1)}%
}\Lambda \ ,\qquad M_{\sigma }\simeq 2\sqrt{6}|C|\left[ \frac{\gamma -1}{2|C|%
}\right] ^{\frac{1}{2(\gamma -1)}}\Lambda \ .
\end{equation}
%We note here
%that the
%quantity $M_{\sigma} <\sigma>$ is invariant under this rescaling procedure.
%Thus
%\begin{equation}
%M_{\sigma} <\sigma>\simeq 2|C| \left[\frac{\gamma -1 }{2|C|} \right]^{\frac{1%
%}{\gamma - 1}} \Lambda ^2 \ .
%\end{equation}
Likewise, in the presence of the quark mass term we have
\begin{equation}
\left[ <\sigma >\right] _{\gamma =1}\simeq \left[ \frac{mN_{f}\Lambda }{2|C|}%
\right] ^{\frac{1}{2}}\ ,\qquad \left[ M_{\sigma }\right] _{\gamma =1}\simeq
2\left[ 2mN_{f}\Lambda \right] ^{\frac{1}{2}}\ .
\end{equation}
Thus the order parameter $\sigma $ for $\gamma =1$ vanishes according to the
power $1/2$ with the quark mass in contrast with an ordinary second order
phase transition where the order parameter is expected to vanish according
to the power $1/3$.

Finally we note an important distinction between our effective potential
describing an infinite order transition and the Ginzburg-Landau potential
describing a second order transition. The latter may be used in both the
symmetric and broken phases, describing light scalar degrees of freedom as
the transition is approached from either side. Our potential develops
infinite curvature at the origin in the symmetric phase, indicating that no
light scalar degrees of freedom are formed as the transition is approached
from that side. This is in agreement with the conclusions of Ref. \cite{ARTW}%
, indicating that as one crosses to the symmetric phase, mesons melt into
quarks and gluons and hence the physics is described via only the underlying
degrees of freedom. The present effective Lagrangian formalism for
describing the chiral/conformal phase transition is close in spirit to the
one developed in Ref.\cite{MY}.

\section{Perturbation Theory and the Determination of $N_f^c$}

Our discussion of the chiral transition so far, using the anomalous
dimension $\gamma$ as the control parameter, has made no direct reference to
$N_f$ and has been independent of perturbation theory. The critical value $%
N_f^c$ for the transition may be estimated by making use of a perturbative
expansion of $\gamma$. Through two orders in perturbation theory, $\gamma$
is given by
\begin{equation}
\gamma =a_0\alpha +a_1\alpha ^2\ ,  \label{gammatot}
\end{equation}
with $\displaystyle{a_0 = \frac1{2\pi} 3C_{2}(R)}$ and $\displaystyle{a_1=
\frac 1{16\pi ^2}\left[ 3C_2(R)^2-\frac{10}3 C_2(R)N_f+\frac{97}3C_2(R)N %
\right] }$, where $\displaystyle{C_2(R)=\frac{N^2-1}{ 2N}}$. We evaluate $%
\gamma $ at the fixed point value of the coupling constant, which at two
loops in the beta function expansion may be expressed as:
\begin{equation}
\alpha ^{*}=-\frac{b_0}{b_1}\simeq \frac{4\pi }N\left[ \frac{11N-2N_f}{
13N_f-34N}\right] \ ,  \label{fixed}
\end{equation}
where we have used the large $N_f$ and $N$ expansion to simplify the
expression.

In Ref.~{\cite{ALM}}, it was noted that the in lowest (ladder) order, the
gap equation leads to the condition $\gamma (2-\gamma )=1$ for chiral
symmetry breaking. To all orders in perturbation theory, this condition is
gauge invariant (since $\gamma $ is gauge invariant) and is equivalent to
the condition $\gamma =1$ Ref.~{\cite{CG}}. To any finite order in
perturbation theory these conditions are of course different. To leading
order in the expansion of $\gamma $, the condition $\gamma (2-\gamma )=1$
leads, to the critical coupling
\begin{equation}
\alpha _{c}=\frac{\pi }{3C_{2}(R)}\ ,  \label{critical}
\end{equation}
above which the ladder gap equation has a non-vanishing solution. Using Eq.~(%
\ref{fixed}) together with Eq.~(\ref{gammatot}) and the condition $\gamma <1$
leads to the conclusion that chiral symmetry is restored for
\begin{equation}
N_{f}>N_{f}^{c}\simeq 3.9N\ .  \label{Ncrit}
\end{equation}
If, on the other hand, the condition $\gamma =1$ is implemented using the
lowest order expression for $\gamma $, a somewhat smaller value of $N_{f}^{c}
$ emerges.

The advantage of using the anomalous dimension $\gamma $ as the control
parameter to study the chiral transition is that the problem can be
formulated in a way that is free of these perturbative uncertainties%
\footnote{%
Recently a perturbative study of the conformal window region in QCD and
supersymmetric QCD was performed in Ref.~{\cite{GG}}}.

\section{Conclusions}

We have explored the chiral phase transition for vector-like $SU(N)$ gauge
theories as a function of the number of flavors $N_f$ via an anomaly induced
effective potential. The effective potential was constructed by saturating
the trace and axial anomalies. It depends on the full beta function and
anomalous dimension of the quark-mass operator. The mesonic degrees of
freedom are the only variables included at low energies. We assumed the
anomaly induced effective potential to have a holomorphic structure. We note
that holomorphicity was also used by other groups \cite{HZ} to constrain
similar anomaly induced potentials. The present potential is a
generalization of a previous potential \cite{toy,HSS,Sannino} which was
constructed by saturating the QCD anomalies at just one-loop.

We showed that the anomaly induced effective potential for $N_{f}>N$ is
positive definite and vanishes with the field $M_{i}^{j}$. We then
investigated the stability of the potential at the origin, and discovered
that the second derivative is positive and divergent when the underlying $%
\beta $ function and the anomalous dimension of the quark-mass operator
satisfy the relation of Eq.~(\ref{criticality}). We took this to be the
signal for chiral restoration. With conformal symmetry being restored along
with chiral symmetry (due to the $\beta $ function vanishing at an infrared
fixed point), the criticality relation becomes a constraint on the anomalous
dimension of the quark-mass operator:
\begin{equation}
\gamma <1\ .
\end{equation}
To convert this inequality into a condition for a critical number of
flavors, we used the perturbative expansion of the anomalous dimension
evaluated at the fixed point, deducing that chiral symmetry is restored for $%
N_{f}\simeq 4N$, in agreement with gap equation studies.

The core of this paper is the proposal that the chiral/conformal phase
transition, suggested by gap equation studies to be continuous and infinite
order, may be described by an effective potential whose form is dictated by
the trace and axial anomalies of the underlying $SU(N)$ gauge theory. It
will be important to explore more completely both the derivation of this
potential and its application to the chiral/conformal phase transition. In
particular we note that we have here used the potential only at the
classical, mean-field level. The development of kinetic energy terms and the
consideration of long wavelength quantum fluctuations of $M_{i}^{j}$ could
next be considered.

\acknowledgments
We are indebted to Thomas Appelquist for enlightening discussions, helpful
comments and for careful reading of the manuscript.  One of us (F.S.) is
happy to thank Gabriele Veneziano for interesting discussions. We are also
happy to thank Amir Fariborz for helpful discussions. The work of F.S. has
been partially supported by the US DOE under contract DE-FG-02-92ER-40704.
The work of J.S. has been supported in part by the US DOE under contract
DE-FG-02-85ER 40231.

\appendix

\section{Higher loop effects for the Effective Potential}

\label{a} The potential in Eq.~(\ref{inter}), when $f(g)$ is evaluated to
lowest order in perturbation theory ($\gamma =0$ and $\displaystyle
{-\frac{\beta (g)}{g^{3}}16\pi ^{2}=\frac{11}{3}N-\frac{2}{3}N_{f}}$), leads
to Eq.~(\ref{hss}). Here we note that the special location of the
singularity in that potential is an artifact of lowest order perturbation
theory. Let us thus investigate the behavior of $f(g)$ to next order. Thus
\begin{equation}
-\frac{\beta (g)}{g^{3}}16\pi ^{2}=b_{0}+b_{1}\alpha ,\quad \quad \gamma
=a_{0}\alpha \ ,  \label{beta}
\end{equation}
with
\begin{equation}
b_{0}=\frac{11}{3}N-\frac{2}{3}N_{f}\ ,\quad b_{1}=\frac{1}{4\pi }\left(
\frac{34}{3}N^{2}-\frac{10}{3}NN_{f}-\frac{N^{2}-1}{N}N_{f}\right) \ ,\quad
a_{0}=\frac{3}{4\pi }\frac{N^{2}-1}{N}\ ,
\end{equation}
which provides
\begin{equation}
f(g)=\frac{11}{3}\left( N-N_{f}\right) +\left[ \frac{34}{3}N^{2}-\frac{10}{3}%
NN_{f}+2\frac{N^{2}-1}{N}N_{f}\right] \frac{\alpha }{4\pi }\ .
\end{equation}
Imposing the equation $f=0$ we find a zero for
\begin{equation}
N_{f}^{s}=N\frac{1+\frac{17\alpha }{22\pi }N}{1+\frac{3\alpha }{22\pi }%
\left( \frac{5N}{3}-\frac{N^{2}-1}{N}\right) }\ ,  \label{ratio}
\end{equation}
which shows that the singularity at $N_{f}=N$ is shifted once higher order
corrections are included.

Whether even this shifted value has any significance depends on the
magnitude of $\alpha $ (the convergence of the expansion). As indicated in
Section III, $\alpha $ can only be guaranteed to be small when $N_{f}$ is
close to, but below $11N/2$, leading to a weak infrared fixed point. This is
a range (well above $N_{f}^{s}$ ) in which $f(g)$ is clearly negative since $%
\beta $ and $\gamma $ are small. We next decrease $N_{f}$ and see whether $%
f(g)$ has a zero in the range of interest. We reduce $N_{f}$ until it
reaches the value $N_{f}^{c}$ ( Eq.~(\ref{Ncrit})) where the infrared fixed
point (Eq.~(\ref{fixed})) has reached the critical coupling (Eq.~(\ref
{critical})). Assuming only that the perturbative expansions leading to
these expressions are roughly accurate (remember that $\alpha $ is never
larger than its fixed point value), $f(g)$ will be negative throughout the $%
N_{f}$ range from $11N/2$ down to $N_{f}^{c}$, the onset of chiral symmetry
breaking. The quantity $N_{f}^{s}$ is well below $N_{f}^{c}$, and
corresponds, probably, to large values of $\alpha $.

To conclude, there is no evidence that $f(g)$ will have changed sign from
negative to positive when $N_f$ is reduced to the critical value $N_f^c$ of
interest here. $N_f^c$ appears to be safely above any possible zeros of $f(g)
$.

This analysis is based on the existence of an infrared fixed point. It is
instructive and reassuring \footnote{%
Although perturbation theory is now probably less reliable.} to observe from
Fig.~1 that, without requiring the existence of any infrared fixed point,
for any value assumed by the quantity $\alpha N$ (at two loops and for large
$N$) the critical number of flavors $N_{f}^{c}$ (defined as the number for
which the exponent in Eq.~(\ref{2nd}) vanishes) is always greater than $%
N_{f}^{s}$. We notice that for a wide range of values of $\alpha N$,$\
N_{f}^{s}/N$ is below the horizontal line $11/2$, above which asymptotic
freedom is lost. It is also clear that there is a region where $N_{f}^{\ast
} $ (the point where the beta function vanishes) is close to $N_{f}^{c}$
(see Fig.~1). In the large $N$ limit we have:
\begin{equation}
\frac{N_{f}^{c}}{N}\simeq \frac{11+ \frac{17}{2\pi }\alpha N}{5+\frac{1}{\pi
} \alpha N}\ ,\qquad \frac{N_{f}^{\ast }}{N} \simeq \frac{11+\frac{17}{2\pi }
\alpha N}{2+\frac{13}{4\pi }\alpha N}\ .
\end{equation}

\begin{figure}[tbp]
$$\mbox{\epsfig{file=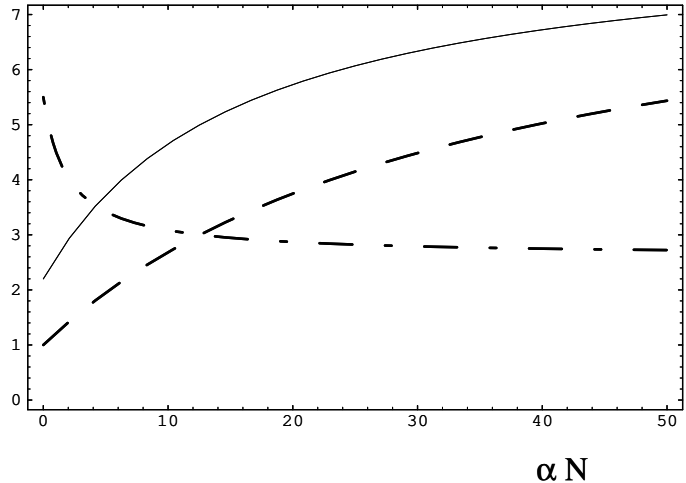,width=9.45cm}} $$
 \caption{$N_{f}^{c}/N$ as a function of $\protect\alpha N$
is shown as a solid line. The dashed line represents
$N_{f}^{s}/N$, while the dot-dashed line describes the
$N_{f}^{\ast }/N$ function.} \label{figura2}
\end{figure}

\end{document}